\documentclass[%
preprint,
 amsmath,amssymb,
 aps,
]{revtex4-1}
\usepackage{comment}
\usepackage{graphicx}% Include figure files
\usepackage{dcolumn}% Align table columns on decimal point
\usepackage{bm}% bold math
\usepackage{natbib}
\begin{document}
\preprint{APS/123-QED}
\title{Weak turbulence in two-dimensional magnetohydrodynamics
%\\with Forced Linebreak
}% Force line breaks with \\
\author{N. Tronko}
\affiliation{Centre for Fusion, Space and Astrophysics, Physics Department, University of Warwick, CV4 7AL, Coventry, UK}
\email{n.tronko@warwick.ac.uk}
\author{S.V. Nazarenko}%
\affiliation{%
 Mathematics Department, University of Warwick, CV4 7AL, Coventry, UK}
\author{S. Galtier}
\affiliation{Institut d'Astrophysique Spatiale, Universit\'e Paris-Sud, \\B\^atiment 121, F-91405, Orsay Cedex, France}
\date{\today}
\begin{abstract}
A weak wave turbulence theory is developed for two-dimensional (2D) magnetohydrodynamics (MHD). We derive and analyze the kinetic equation describing
the three-wave interactions of pseudo-Alfv\'en waves. Our analysis is greatly helped by the fortunate fact that in 2D the wave-kinetic equation is integrable.
In contrast with the 3D case, in 2D the wave interactions are nonlocal. Another distinct feature is that strong derivatives of spectra tend to appear in the region
of small parallel (i.e. along the uniform magnetic field direction) wavenumbers leading to a breakdown of the weak turbulence description in this region.
We develop  a qualitative theory beyond weak turbulence describing subsequent evolution and formation of a steady state.
\begin{description}
\item[PACS numbers] $47.27$ E, $47.65$-d, $52.30$ Cv, $52.35$ Bj, $95.30$ Qd.
\end{description}
\end{abstract}
\pacs{Valid PACS appear here}% PACS, the Physics and Astronomy
\maketitle
%\tableofcontents
\section{Introduction\label{sec:intro}}
Magneto-Hydro-Dynamics (MHD) is of great interest for modeling turbulence in magnetically confined  and unconfined plasmas.
In astrophysics its applications range from solar wind \cite{Marsch_Tu},  to the Sun \cite{Priest}, to the interstellar medium \cite{Ng_Galtier} and
beyond \cite{Zweibel_Heiles}. At the same time, MHD is also relevant to large-scale motion in nuclear fusion devices such as tokamaks \cite{Strauss_1976}.
One of the pioneering results of incompressible MHD turbulence has been obtained by Iroshnikov \cite{Iroshnikov_1963} and Kraichnan \cite{Kraichnan_1965} (thereafter IK)
who proposed an extension of  the Kolmogorov phenomenology \cite{Kolmogorov}, originally derived for Navier-Stokes equations for  Hydro-Dynamics (HD). For simplicity the assumptions of homogeneity and isotropy were made by Kolmogorov, and the energy cascade was supposed to be dominated by local (in scale) interactions
between eddies of similar size. Then the Kolmogorov  phenomenology leads to the well-known one-dimensional kinetic energy spectrum $E(k)\sim k^{-5/3}$, where $k$
is the wave vector. The associated cascade properties for its inviscid invariants differ for 3D and 2D turbulence: while in 3D the energy and the kinetic helicity
exhibit direct cascades, in 2D the energy cascades inversely -- still with a $-5/3$ scaling -- whereas a direct cascade is found for the enstrophy which leads to  spectrum  $E(k)\sim k^{-3}$ at the small scales.

IK modified the Kolmogorov phenomenology by taking into account the magnetic field. They also assumed  turbulence homogeneity, isotropy and locality of interactions.
However, there exist fundamental differences between the Kolmogorov and the IK theories. First of all, in MHD the energy cascade
is supposed to be dominated not by the interactions between eddies but between Alfv\'en wave packets propagating in  opposite directions; this modification
leads to the energy spectrum $E(k) \sim k^{-3/2}$. Furthermore,
unlike hydrodynamics the cascades of the ideal MHD invariants exhibit some similarities in their behaviour in 2D and 3D \cite{Biskamp03}: in both
cases, the cascades have the same direction with a direct cascade for the energy and cross-helicity, and an inverse cascade for the magnetic helicity or the
anastrophy.

The differences between MHD and HD turbulence go beyond these classical properties. In the IK theory the large-scale magnetic field is supposed
to play the role of external field which is necessary for the existence of Alfv\'en waves but its main effect, i.e. anisotropy, is not taken into account. The importance
of an external magnetic field has been discussed many times during the last two decades \cite{Montgomery_Turner,Shebalin_Matthaeus,Oughton_Priest,Ng_Bhattacharjee,Bigot_Galtier_PRL,Verma_2004,Chandran_2005,Boldyrev_2006,Bigot_PRE}
and the anisotropic behavior has been shown in direct numerical simulations for both 2D \cite{Shebalin_Matthaeus} and 3D \cite{Oughton_Priest}.

Despite some similarities -- like the cascade directions -- the question about the identification of differences between 2D and 3D MHD turbulence still
represents an important issue. In early numerical studies \cite{Shebalin_Matthaeus} mainly 2D simulations were performed because of the limited numerical
resources available and the non-accessibility of 3D calculations at high Reynolds numbers. Nowadays, 3D MHD numerical simulations are commonly achieved
\cite{Politano_Pouquet_lett,Gomez_Politano,Biskamp_Mueller,Haugen_Bradenburg,Biskamp_Schwartz,Biskamp_Mueller,Merrifield_Chapman,Merrifield_2,Bigot_Galtier_PRE}. At the same time, 2D simulations are still  used for the illustration of new numerical techniques \cite{Dritchel_Tobias_2012}. There is also an interest for
the investigation of freely decaying MHD turbulence because the Reynolds numbers can be higher in 2D than in 3D \cite{Biskamp_Schwartz}.
When the magnetic field perturbations are small compare to a uniform background magnetic field, the 2D MHD equations are sometimes used to model turbulence.
Such a situation is particularly relevant for solar coronal loops as well as for reduced models for fusion plasmas in slab geometry \cite{Strauss_1976}.

Strong statements were made by some authors that 2D simulations can be safely used to model 3D situations because the properties of the 2D and the 3D MHD turbulence are essentially the same \cite{Biskamp_Schwartz,Biskamp}. One of the motivations of the present paper is to test validity of this claim in a special case when the external magnetic field is strong.

In this paper, we consider 2D MHD in the presence of a strong background magnetic field which implies realization of the weak turbulence regime.
One of the main advantages of this regime is the fact that it allows one to derive accurate analytical results for the spectrum. An explicit comparison will be made
between the weak turbulence regimes in 2D and in 3D; the latter was analyzed rigorously in \cite{Galtier_2000}.
The weak wave turbulence approach is widely familiar to the plasma physics community \cite{Achterberg_1979,Akhiezer_1975,McIvor_1977,
Zakharov_1974,Zakharov_1984,Zakharov_1992,Kuznetsov_1972,
Kuznetsov_1973,Kuznetsov_1973,Tsytovich,Sagdeev_Galeev,Vedenov,Galtier_2001,Galtier_2008}. It is a statistical description of a large ensemble of weakly interacting dispersive waves. The formalism leads to wave kinetic equations from
which exact power law solutions can be found for the energy spectra.
There were several reasons which postponed development of weak turbulence for Alfv\'en waves. The first one is their semi-dispersive nature. Typically, the
wave-kinetic approach cannot be used for non-dispersive waves since such wavepackets propagate with the same group velocity even if their wavenumbers are
different, the energy exchange between such waves may not be considered small, and may lead to possible energy accumulation over a long time of interaction.
The Alfv\'en waves represent a unique exception to this rule because co-propagating waves do not interact and the nonlinear interaction is present only for counter-propagating waves. The latter pass through each other in some finite time and no long time cumulative effect  occur.
That is why the Alfv\'en waves represent a unique example of semi-dispersive waves for which the wave turbulence theory applies.
The second reason which renders the weak turbulence theory for Alfv\'en waves very subtle is the fact that the domination of three-wave interactions --
as assumed by IK -- may be questionable. While in \cite{Shridar_Goldreich} the three-wave interactions were declared absent, the IK argument has been
re-established in \cite{Galtier_2000,Ng_Bhattacharjee,Montgomery_Matthaeus}.
The weak turbulence theory for 3D incompressible MHD was developed in \cite{Galtier_2000} (see also \cite{Galtier_2002,Nazarenko_book}) where
three-wave kinetic equations were derived with their exact solutions {\it via} a systematic asymptotic expansion in powers of small nonlinearities.

The main goal of the present paper is to derive the weak turbulence equation for the 2D MHD, analyze it and make a comparison with the 3D case.
The crucial technical step which allows a comprehensive theoretical analysis of the solutions consists in transforming the wave kinetic equation into an integrable
form by Fourier transforming it and separating the transverse and the parallel dynamics by using a self-similar ``effective time" variable.

This article is organized as follows.
In section \ref{sec:wave_kin_eqn} we derive the weak turbulence kinetic equation through a general perturbative procedure. In sections \ref{sec:kz_spectra} and \ref{sec:kin_eqn_dyn} we proceed with a detailed investigation of its properties in an anisotropic limit. The main goal of such a study is to verify whether or not the
turbulence is local. First of all, we will consider the steady state behavior by looking for Kolmogorov-Zakharov type solutions and check their locality.
Next, we will proceed with investigation of an unsteady spectrum evolution by considering two different cases with a gaussian-shaped source and different kinds of
dissipation: a uniform friction and a viscosity. Due to integrability of the weak wave-kinetic equation in the first case it is possible to find an exact solution. In  the
second case, a qualitative analysis for the steady state is complemented by a numerical simulation of the spectrum evolution.
The goal of section \ref{sec:beyond_wt} is to develop some qualitative reasoning about the turbulent behavior of our system near the applicability margin of
wave-kinetic formalism  and  beyond. Formation of steady state is also discussed. Finally, we present a summary of our results in section \ref{sec:conclusions}.
%%%%%%%%%%%%%%%%%%%%%%%%%%%%%%%%%%%%%%%%%%%%%%%%%%%%

%%%%%%%%%%%%%%%%%%%%%%%%%%%%%%%%%%%%%%%%%%%%%%%%%%%%%%%%%%%
%
%
\section{Wave-kinetic description}
\label{sec:wave_kin_eqn}
%
%%%%%%%%%%%%%%%%%%%%%%%%%%%%%%%%%%%%%%%%%%%%%%%%%%%%%%%%%%%

%%%%%%%%%%%%%%%%%%%%%%%%%%%%%%%%%%%%%%%%%%%%%%%%%%%
%
\subsection{Alfv\'en waves}
\label{sec:alfven_waves}
%%%%%%%%%%%%%%%%%%%%%%%%%%%%%%%%%%%%%%%%%%%%%%%%%%%
In 3D incompressible MHD there exist two different kinds of Alfv\'en waves \cite{Verma_2004}.
The first kind, called Shear-Alfv\'en waves (SAW), have fluctuations of velocity and magnetic field transverse to the initial background magnetic field $\mathbf{B}_0$ , whilst the other kind, called Pseudo-Alfv\'en waves (PAW), have fluctuations along $\mathbf{B}_0$. Both waves propagate along $\mathbf{B}_0$ at the same Alfv\'en speed.

The weak wave turbulence approach for incompressible MHD applies for a small nonlinearity, $\epsilon \sim
b_{\perp} k_{\perp}/ B_0 k_{||} \ll 1$, were $b_{\perp}$ is the perpendicular magnetic field perturbation
 and $k_{||}$ and $k_{\perp}$ are the wave vector components in the parallel and the perpendicular directions to $\mathbf{B}_0$. Additionally,
 a strong anisotropy condition is often used,   $\sigma =k_{\perp}/k_{||} \gg 1$.
In the 3D case, it was shown that in the leading order of weak nonlinearity,  $\epsilon \ll 1$,  and strong anisotropy,  $\sigma \gg 1$, the SAW interact only among themselves and evolve independently from the PAW. At the same time, the PAW scatter from the SAW without amplification or damping, and they do not interact with each other. Such a behaviour does not rule out a possibility  for the PAW to be interacting among themselves in the next order of expansion in $1/\sigma $. However in the 3D case, such a process is sub-dominant to a stronger interaction with the SAW and was not considered yet.

In the 2D case, due to the geometrical restrictions, it is only possible to have the PAW and not SAW.
In this paper we will see that three-wave interactions of PAW do occur in the 2D case in the next order of expansion in $1/\sigma $ and represent the dominant process in the nonlinear evolution.

%%%%%%%%%%%%%%%%%%%%%%%%%%%%%%%%%%%%%%%%%%%%%%%%%%%%%%%
%
\subsection{Interaction representation.}
%
%%%%%%%%%%%%%%%%%%%%%%%%%%%%%%%%%%%%%%%%%%%%%%%%%%%%%%%
The ideal incompressible MHD system in Els\"asser  variables $\mathbf{z}=\mathbf{v}+s\mathbf{b}$, with $s=\pm 1$, is given by \cite{Biskamp}
\begin{eqnarray}
\left(\partial_t-s{\mathbf B}_0\cdot\nabla+
\mathbf{z}^{-s}\cdot\nabla\right)&{\mathbf z}^s=&-\nabla P_{*},\\
&\nabla\cdot\mathbf{z}^s&=0,
\label{eq:mhd_free}
\end{eqnarray}
where $\mathbf{v}$ is the fluid velocity, $\mathbf{b}$ is the magnetic field fluctuation (in velocity units), $\mathbf{B}_0$ is a uniform background magnetic field
(also  in velocity units, i.e. the Alfv\'en speed) and $P_*$ is the total (thermal plus magnetic) pressure.
In what follows we suppose that the background magnetic field is directed along the $\widehat{\mathsf x}$ axis,
${\mathbf B}_0=B_0\widehat{\mathsf x}$.
In the coordinate notations we have
\begin{eqnarray}
\left(\partial_t-s B_0 \partial_x\right)z_j^s=-\epsilon\ z_n^{-s}\partial_n
z_j^{s}-\partial_j P_*,
\label{eq:mhd_1}
\\
\partial_i z_i^s=0.
\label{eq:mhd_2}
\end{eqnarray}
The nonlinear terms in eq.(\ref{eq:mhd_1}) include  Els\"asser variables of the opposite signs only.
Therefore, the nonlinear interactions take place only between counter-propagating waves.

The first step in the general procedure of the wave kinetic formalism  is to identify the linear modes. Neglecting the nonlinear  terms in the right hand side (r.h.s.) of the equation (\ref{eq:mhd_1}) (which includes the pressure term) and  looking for solutions in the form of a wave,
\begin{equation}
z^s_j\sim e^{i(k_x x+k_y y)-i\omega^s t},
\end{equation}
we get two linear modes,
\begin{equation}
\omega^s=- s B_0 k_x, s=\pm 1,
\end{equation}
which propagate parallel to the background magnetic field (in both directions) with group velocity $\mathbf{v}_g^s=- s \mathbf{B}_0$. Let us suppose that our
system is periodic in the physical space with period $L$ in both $x$ and $y$,
and let us introduce the Fourier series:
\begin{equation}
z_j^s({\mathbf x},t)=\sum_{{\bf k}}  a_j^s({\mathbf k},t) e^{i {\mathbf k}\cdot{\mathbf x}},
\end{equation}
where wave vector ${\bf k}$ takes values on a 2D grid, ${\bf k}=(k_x,k_y) =
(2 \pi m_x/L, 2 \pi m_y/L)$ where $m_x,m_y \in Z$.
Then, by applying the divergence operation on both sides of the equation (\ref{eq:mhd_1}) and by using (\ref{eq:mhd_2}), we find the expression for the Fourier
coefficients of pressure $P_*$:
\begin{equation}
\hat P_{*}(\mathbf{k})=-k^{-2} \sum_{{\bf k}_1, {\bf k}_2} (\mathbf{k}_2\cdot\mathbf{a}^{-s}(\mathbf {k}_1,t))(\mathbf{k}\cdot\mathbf{a}^s(\mathbf {k}_2,t)) \delta(\mathbf{k}_1+\mathbf{k}_2-\mathbf{k}),
\end{equation}
where $\delta(\mathbf{p}) $ is the Kronecker delta: $\delta(\mathbf{p})=1 $ for $\mathbf{p}=0 $ and zero otherwise. Thus, equation (\ref{eq:mhd_1}) in Fourier space
becomes:
\begin{equation}
%\float
(i\partial_t -\omega^s) \mathbf{a}^s(\mathbf{k},t)=
\hspace{-.1cm}
\sum_{{\bf k}_1, {\bf k}_2}  (\mathbf{k}\cdot\mathbf{a}^{-s}(\mathbf {k}_1,t))
\left[ \mathbf{a}^s(\mathbf {k}_2,t)-\frac{\mathbf{k}} {k^2}
(\mathbf{k}\cdot\mathbf{a}^s(\mathbf {k}_2,t))\right]
\hspace{-.1cm}
\delta(\mathbf{k}_1+\mathbf{k}_2-\mathbf{k}) .
\label{eq:mhd_fourier}
\end{equation}
Using the incompressibility condition
\begin{equation}
a_x^s=-a_y^s\frac{k_y}{k_x} \, ,
\end{equation}
we reduce (\ref{eq:mhd_fourier}) to one scalar equation,
\begin{equation}
%\ltx
(i\partial_t -\omega^s) a_x^s(\mathbf{k},t)=\epsilon \sum_{{\bf k}_1, {\bf k}_2}\int
k_y\ \frac{(\mathbf{k}\times\mathbf{k}_1)_z \left(\mathbf{k}\cdot\mathbf{k}_2\right)}{k_{1y}\ k_{2y}\ k^2}\
a_x^{-s}(\mathbf{k}_1,t)\ a_x^s(\mathbf{k}_2,t)
d {\mathbf k}_1\ d {\mathbf k}_2.
\end{equation}
Let us now introduce the representation of interaction variables,
\begin{equation}
c_{\bf k}^s=  i\frac{k}{k_y\epsilon} a_x^s(\mathbf{k},t) e^{i\omega^s t},
\end{equation}
which represent slowly varying wave amplitudes. Factor $e^{i\omega^s t}$ here compensates for the fast-scale  oscillations arising due to the linear dynamics. We have introduced a formal constant small parameter $\epsilon\ll 1$ for easier counting of the powers of nonlinearity, assuming now that $c_{\bf k}\sim 1$.
Then the system (\ref{eq:mhd_fourier}) in the interaction representation variables is:
\begin{equation}
\partial_t c_{\bf k}^{\pm}= \epsilon \sum_{{\bf k}_1, {\bf k}_2} V_{1 2 k}\
c^{\mp}_{{\bf k}_1}\ c_{{\bf k}_2}^{\pm} e^{2 ik_{1x} t} \delta({\mathbf k}_1+{\mathbf k}_2-{\mathbf k}),
\label{eq:mhd_interact}
\end{equation}
with the interaction coefficient:
\begin{equation}
V_{1 2 k}=\frac{\left({\mathbf k}\cdot {\mathbf k}_2\right)\left[{\mathbf k_1}\times{\mathbf k_2}\right]_z}{k\ k_1\ k_2}.
\label{int_kernel}
\end{equation}
Note that so far we have not used smallness of $\epsilon$ and the eq.(\ref{eq:mhd_interact}) is completely equivalent to the initial system (\ref{eq:mhd_1})-(\ref{eq:mhd_2}).

%%%%%%%%%%%%%%%%%%%%%%%%%%%%%%%%%%%%%%%%%%%%%%%%%%%%%%%%%%%%
%
\subsection{Wave-kinetic equation}
%
%%%%%%%%%%%%%%%%%%%%%%%%%%%%%%%%%%%%%%%%%%%%%%%%%%%%%%%%%%%%
The standard weak turbulence approach \cite{Zakharov_1992,Nazarenko_book} exploits the smallness of nonlinearity, randomness of phases and the infinite
box limit. In Appendix \ref{app:WT_procedure}, we are applying this approach to the system (\ref{eq:mhd_interact}) and  obtain the following kinetic equation for the
wave spectrum ${n}_{\bf k}$,
\begin{equation}
\dot{n}_{\bf k}^{\pm}=\pi  \int V_{k 12}^2 n^{\mp}_{{\bf k}_1}\left[ n_{{\bf k}_2}^{\pm}- n_{\bf k}^{\pm}\right]
\delta\left(\mathbf{k}-\mathbf{k}_1-
\mathbf{k}_2\right)\delta(2 k_{1x}) d\mathbf{k}_1\ d\mathbf{k}_2 \, ,
\label{kin_eq}
\end{equation}
where the interaction coefficient is as in expression (\ref{int_kernel}). Here, $\delta({\bf p})$ means Dirac's delta function.
In the following sections we proceed with detailed analysis of this equation.
%%%%%%%%%%%%%%%%%%%%%%%%%%%%%%%%%%%%%%%%%%%%%%%%%%%%%%%%%%%%%%
%
\subsection{Anisotropic limit}
%
%%%%%%%%%%%%%%%%%%%%%%%%%%%%%%%%%%%%%%%%%%%%%%%%%%%%%%%%%%%%%%
One remarkable property of MHD turbulence, which makes it very different from the HD one, is its strong anisotropy in presence of strong background
magnetic field. It was illustrated with direct numerical simulations in both 2D \cite{Shebalin_Matthaeus} and 3D \cite{Oughton_Priest}.
The wave-kinetic formalism confirms such an anisotropy through the  kinetic equation structure.
In fact for Alfv\'en waves, the resonant three-wave interaction  \cite{Shebalin_Matthaeus} is organised in such a way that one member of each triad must have its wave vector perpendicular to the external magnetic field. At the same time, the two other waves in the same triad must have their parallel wavenumbers equal to each other: $k_{||}=k_{2||}$. Formally, this is seen in both the 2D and the 3D kinetic equations whose r.h.s.  contains a delta function $\delta(2 k_{1 ||})$, see equation (\ref{kin_eq})
for the 2D case and the equation (26) in \cite{Galtier_2000} for the 3D case. Using this delta function and integrating over $k_{1 ||}$
we see that the parallel component of the wavenumber enters into the kinetic equation as an external parameter and the spectrum dynamics is decoupled at each
level of $k_{ ||}$. In other words, there is no energy transfer in the parallel (to the external field ${\bf B_0}$) direction in the $\mathbf k$-space.
The initial  spectrum is spreading over the transverse wavenumbers $k_{\perp}$, and not in the $k_{||}$ direction. For  large time, such a spectrum becomes very flat,  pancake-like.

The two-dimensionalisation of the total energy means that, for large time, the energy spectrum is supported on a volume of wave-numbers such that for
most of them  $k_{\perp} \gg k_{||}$.

Thus, let us consider the anisotropic limit for kinetic equation (\ref{kin_eq}), which reads in 2D as $k_y\gg k_x$.
Taking into the account the resonant interaction conditions for the parallel wavenumbers, we will have
a considerable simplification of the interaction coefficient:
\begin{equation}
 V_{k 12}=k_{x},
\end{equation}
and the kinetic equation will take the following form,
\begin{eqnarray}
%\fl
\dot{n}^{\pm}(k_x,k_y)=\pi   k_x^2 \int  n^{\mp}(0,k_{1 y}) [ n^{\pm}(k_{x}, k_{2 y})- n^{\pm}(k_x,k_y) ]
\label{eq:kin_anisotropy}
%\hspace{-.1cm}
\delta(k_y-k_{1 y}-
k_{2 y}) d k_{1 y} d k_{2 y}.
\end{eqnarray}
This equation describes three-wave interactions of PAW in 2D in the anisotropic limit.
One can immediately see that the energy is conserved separately in the
"+" and "-" waves separately at each $k_x$:
\begin{equation}
\partial_t  \int  n^{\mp}(k_x, k_y,t)
\label{eq:ener}
%\hspace{-.1cm}
 d k_{y}  = 0.
\end{equation}

Factor $k_x^2$ in the r.h.s. of  relation (\ref{eq:kin_anisotropy}) is very important.
In the 3D case, there exists a similar term which corresponds to a sub-leading contribution. We remind that in 3D in the leading order of the perturbation theory,
the PAW are scattered on SAW and do not interact directly to each other. In the 2D, there is no SAW and, therefore, the
r.h.s.  of (\ref{eq:kin_anisotropy}) becomes the leading order contribution.

Further, in leading order of 3D there is no $k_x^2$ factor, and substitution $n(k_\perp, k_{||}, t) =
n_\perp(k_\perp, t)n_{||}(k_{||})$ leads to an equation for $n_\perp(k_\perp, t)$ which does not involve $k_{||}$.
In 2D, one can also obtain an equation which does not involve $k_x$ but for this, one has to introduce an ``effective time" variable
 $\tau=\pi k_x^2 t$:
\begin{eqnarray}
%\fl
\partial_{\tau}{n}^{\pm}(k_{x},k_y;\tau)= \int  n^{\mp}(0,k_{1y};0)\left[ n^{\pm}(k_{x},k_{2y};\tau)- n^{\pm}(k_{x},k_y;\tau)\right]
\nonumber\\
\hspace{6cm}
\times\delta\left(k_y-k_{1y}-k_{2y}\right) d k_{1y} dk_{2y} \, .
\label{kin_eqn_simpl}
\end{eqnarray}
Now we can seek solution in the following form,
\begin{eqnarray}
n(k_x,k_y,\tau)=\mu(k_x)\eta(k_y,\tau),
\label{mueta}
\end{eqnarray}
where $\mu(k_x)$ represents the parallel (non-evolving) component of the energy spectrum and  $\eta(k_y,\tau)$ is the perpendicular one.
Without loss of generality we can assume $\mu(0)=1$. Substituting  (\ref{mueta}) into (\ref{kin_eqn_simpl}) we have the following equation for $\eta$,
\begin{eqnarray}
%\fl
\partial_{\tau}{\eta}^{\pm}(k_y,\tau)= \int  \eta^{\mp}(k_{1y},0)\left[ \eta^{\pm}(k_{2y},\tau)- \eta^{\pm}(k_y,\tau)\right]
\delta\left(k_y-k_{1y}-k_{2y}\right) d k_{1y} dk_{2y} \, .
\label{kin_eta}
\end{eqnarray}
Thus, like in 3D, we have an evolution equation for the perpendicular part of the spectrum which does not explicitly depend on the parallel
part. However, the qualitative difference with the 2D case  is that there is an implicit dependence on $k_{x}$  via the effective time variable $\tau$,
which in particular leads to the fact that in the r.h.s. of equation (\ref{kin_eta}) one of $\eta$'s is taken at $\tau =0$, making this
equation linear and, as we will soon see, integrable. Another distinct feature arising from such an implicit dependence
on $k_{x}$  via  $\tau$ is sharpening of the spectrum at small $k_{x}$
leading to the breakdown of the wave-kinetic description. Later in this paper we will study this effect and its consequences.

\section{Kolmogorov-Zakharov spectra and their locality}
\label{sec:kz_spectra}
%
%
%%%%%%%%%%%%%%%%%%%%%%%%%%%%%%%%%%%%%%%%%%%%%%%%%%%%%%%%%%%%%%%%
At the first step of our investigation of the wave-kinetic equation (\ref{kin_eta}), we seek  exact stationary power law solutions, $ \eta(k_y;\infty)^{\pm}\propto k_y^{\nu^{\pm}}$. For this, we will use so-called Zakharov transformation,
\begin{equation}
k_{1 y}'=\frac{k_y k_{1y}}{k_{2y}},\ k_{2y}'=\frac{k_y^2}{k_{2y}} \,.
\label{Zakharov_transform}
\end{equation}

We split the integral into the r.h.s. of (\ref{kin_eta}) in two equal parts and we perform the Zakharov  transformation in the integrand of one of these parts.
This leads to  the following conditions on the exponents,
\begin{equation}
\nu^++\nu^-=-2 \, .
\end{equation}
However,  the resulting power law spectra represent true mathematical solutions if and only if the
original (before the transformation) integral in the r.h.s. of (\ref{kin_eta}) converges at these solutions.
In Appendix \ref{app:convergence}, we perform such a convergence study and show that this integral never converges and therefore, no Kolmogorov-Zakharov
solutions are possible for 2D MHD system, contrary to the 3D case. We remark that the balanced turbulence spectrum with $\nu^+=\nu^-=-1$ has a logarithmic
divergence. Thus, one could anticipate that such a  marginal nonlocality could be ``fixed" by
logarithmic corrections. We will see later that this is false.

%%%%%%%%%%%%%%%%%%%%%%%%%%%%%%%%%%%%%%%%%%%%%%%%%%%%%%%%%%%%%%%%
%
%
\section{Integration of  the kinetic equation}
\label{sec:kin_eqn_dyn}
%
%
%%%%%%%%%%%%%%%%%%%%%%%%%%%%%%%%%%%%%%%%%%%%%%%%%%%%%%%%%%%%%%%%
A remarkable property of the kinetic equation (\ref{kin_eqn_simpl}) is its simplicity.
In this section we show that in some physical situations it can be solved analytically.
Let us introduce into the equation (\ref{kin_eqn_simpl}) sources and sinks of waves:
\begin{eqnarray}
%\fl
\partial_{\tau}{n}^{\pm}(k_{x},k_y,\tau)=  \int  n^{\mp}(0,k_{1y},0)\left[ n^{\pm}(k_{x},k_{2y},\tau)- n^{\pm}(k_{x},k_y,\tau)\right]
\nonumber\\
\hspace{1cm}
\times\delta\left(k_y-k_{1y}-k_{2y}\right) d k_{1y} dk_{2y}
+ {\mathcal F}(k_y,k_x)-\sigma_d\ n(k_x,k_y,\tau).
\end{eqnarray}
The function ${\mathcal F}(k_x,k_y)$ may represent forcing or dissipation, depending on the choice of the sign before it, and the constant $\sigma_d$ introduces a uniform friction.

In order to use factorisation (\ref{mueta}) and eliminate $\mu(k_x)$ in both sides of the forced kinetic equation, we assume  the following type of force/dissipation function, ${\mathcal F}(k_y,k_x)={\mathcal F}_x(k_x){\mathcal F}_y(k_y)$. Then  the parallel component of (\ref{mueta})
must be chosen as $\mu(k_x) = {\mathcal F}_x(k_x)$.
Finally, we obtain the following forced/dissipated kinetic equation for the perpendicular component of the energy spectrum,
\begin{eqnarray}
%\fl
\partial_{\tau}{\eta}^{\pm}(k_y,\tau)=  \int  \eta^{\mp}(k_{1y},0)\left[ \eta^{\pm}(k_{2y},\tau)- \eta^{\pm}(k_y,\tau)\right]
\nonumber\\
\hspace{2cm}
\times\delta\left(k_y-k_{1y}-k_{2y}\right) d k_{1y} dk_{2y} + {\mathcal F}_y (k_y)-\sigma_d \eta(k_y,\tau) \, .
\label{kin_eqn_perp_forced}
\end{eqnarray}
%%%%%%%%%%%%%%%%%%%%%%%%%%%%%%%%%%%%%%%%%%%%%%%%%%%%%%%%%%%%%%%%
%
\subsection{Pseudo-physical space}
%
%%%%%%%%%%%%%%%%%%%%%%%%%%%%%%%%%%%%%%%%%%%%%%%%%%%%%%%%%%%%%%%%
A considerable simplification of  equation (\ref{kin_eqn_perp_forced}) can be obtained with performing the inverse Fourier transform on $\eta(k_y,\tau)$:
\begin{equation}
\mathcal{E}^{\pm}\left(y,\tau\right)=\int\eta^{\pm}\left(k_y,\tau\right) e^{i k_y y} dk_y.
\end{equation}
We will call $\mathcal{E}^{\pm}\left(y,\tau\right)$ the pseudo-physical space energy, keeping in mind that what is transformed is the spectrum and not the original wave variable.

We arrive at the following representation of equation (\ref{kin_eqn_perp_forced}) in the pseudo-physical space,
\begin{equation}
\partial_{\tau}\mathcal{E}^{\pm}\left(y,\tau\right)= \mathcal{E}^{\mp}\left(y,\tau\right)
\left[\mathcal{E}^{\pm}\left(y,0\right)-\mathcal{E}^{\pm}\left(0,0\right)-\sigma_d\right]+\widehat{{\mathcal F}}(y).
\label{pseudo_fourier}
\end{equation}

%%%%%%%%%%%%%%%%%%%%%%%%%%%%%%%%%%%%%%%%%%%%%%%%%%%%%%%%%%%%%%%%
%
\subsection{General solutions}
%
%%%%%%%%%%%%%%%%%%%%%%%%%%%%%%%%%%%%%%%%%%%%%%%%%%%%%%%%%%%%%%%%
Let us  consider the balanced turbulence case with $\mathcal{E}^+(y,\tau)=\mathcal{E}^-(y,\tau)$.
Then, the general solution of equation (\ref{pseudo_fourier}) can be written as:
\begin{equation}
\mathcal{E}(y,\tau)=C(y) e^{\left(\mathcal{E}(y,0)-\mathcal{E}(0,0)-\sigma_d\right)\tau}-\frac{\widehat{{\mathcal F}}(y)}{\mathcal{E}(y,0)-\mathcal{E}(0,0)-\sigma_d},
\label{an_gen}
\end{equation}
where the first term represents the general solution for the homogeneous equation and the second term is a particular (time independent) solution 
 of the inhomogeneous equation.
Function $C(y)$ has to be fixed by the initial condition,
\begin{equation}
C(y)= \mathcal{E}(y,0)+\frac{\widehat{{\mathcal F}}(y)}{\mathcal{E}(y,0)-\mathcal{E}(0,0)-\sigma_d}.
\label{const}
\end{equation}

Now let us consider two particular examples of the forcing and dissipation.  In both cases we will assume a Gaussian shape forcing,
$ \widehat{{\mathcal F}}(y)  = \sigma_f e^{k_f ^2 y^2/2}$, where constants $\sigma_f $ and $k_f $ represent
the forcing strength and its characteristic wave vector respectively (in the ${k_y}$-space the forcing is also Gaussian, centered at
${k_y}=0$ and with width $k_f $).
 In the first example, the dissipation will be represented by a uniform friction. Here, we can write analytical solutions of the kinetic equation on
 the pseudo-physical space. In the second case, we will consider a viscous dissipation.
Here, a qualitative analysis of   the stationary regime can be done. In order to illustrate the spectrum evolution in that case, a numerical solution will be used.

%%%%%%%%%%%%%%%%%%%%%%%%%%%%%%%%%%%%%%%%%%%%%%%%%%%%%%%%%%%%%%%%
%
\subsubsection{Uniform friction.}
%
%%%%%%%%%%%%%%%%%%%%%%%%%%%%%%%%%%%%%%%%%%%%%%%%%%%%%%%%%%%%%%%%

For  the uniform friction case we have:

\begin{equation}
\mathcal{E}\left(y,\tau\right)=C(y) e^{\left(\mathcal{E}(y,0)-\mathcal{E}(0,0)-\sigma_d\right)\tau}-
\frac{\sigma_f e^{k_f^2 y^2/2}}{\mathcal{E}(y,0)-\mathcal{E}(0,0)-\sigma_d}.
\end{equation}
For simplicity, let us use  a single-wave initial condition
\begin{equation}
\mathcal{E}(y,0)=2 A \cos\left(k_0 y\right), \;\;\; A=\hbox{const} >0,
\label{initial}
\end{equation}
 which corresponds to two $\delta$-functions in $k_y$-space: at $k_y = \pm k_0$.

Then, we can find a function $C(y)$ using equation (\ref{const}) and substitute it into our solution, which yields:
\begin{eqnarray}
\nonumber
%\fl
\mathcal{E}(y,\tau) =\left[2 A \cos(k_0 y)+\frac{\sigma_f e^{{-k_f^2 y^2}/{2}}}{2 A \left(\cos(k_0 y)-1\right)-\sigma_d}\right]
e^{\left(2A\left(\cos k_0 y-1\right)-\sigma_d\right)\tau}\\
\hspace{5cm}
-
\frac{\sigma_f e^{{-k_f^2 y^2}/{2}}}{2 A \left(\cos(k_0 y)-1\right)-\sigma_d}.
\label{general_solution}
\end{eqnarray}

Let us examine the steady state, which corresponds to the limit $t\rightarrow\infty$ and, therefore,
 $\tau\rightarrow\infty$. Note, however, that the time for the steady state to form becomes longer as $k_x$ gets less, and there are always
very small $k_x$ where the spectrum is evolving at any large time.
In the limit $\tau\rightarrow\infty$ the solution is given by the second term in the r.h.s. of
(\ref{general_solution}). Far from the initial and the forcing scales, at $k\gg k_0$ and $k\gg k_f$, which
corresponds to $y\ll 1/k_0$ and $y\ll 1/ k_f$, we have
 $\cos(k_0 y)=1-(k_0 y)^2/2+\mathcal{O}((k_0 y)^4)$ and $e^{{-k_f^2 y^2}/{2}} = 1+ \mathcal{O}((k_f y)^2)$.
Thus for this range of scales we have the following expression for the steady state solution in the pseudo-Fourier space,
\begin{equation}
\mathcal{E}\left(y,\infty\right)=\frac{\sigma_f}{\sigma_d+\lambda y^2}, \;\;\;\; \hbox{where} \;\;\;\; \lambda=A k_0^2.
\label{steady_pseudo}
\end{equation}
 Performing  Fourier transform of $\mathcal{E}(y,\infty)$ we  get the steady spectrum,
\begin{equation}
\eta\left(k_y,\infty\right)=\frac{1}{2\pi}\int_{-\infty}^{\infty}\mathcal{E}\left(y,\infty\right)e^{-i k_y y} dy.
\label{inverse_fourier}
\end{equation}
For  wavenumbers in the inertial range $k_0,k_f \ll k\ll k_d = \sqrt{\lambda/\sigma_d}$, expression (\ref{steady_pseudo})
becomes effectively a delta function in the integrand of (\ref{inverse_fourier}),
\begin{equation}
\frac{\sigma_f}{\sigma_d+\lambda y^2}
 \approx
\frac{\pi \sigma_f }{ \sqrt{\sigma_d\lambda}}\delta({y}),
\label{delta-lorent}
\end{equation}
and
we have
\begin{equation}
\eta\left(k_y;\infty\right)=\frac{1}{2}\frac{\sigma_f}{\sqrt{\sigma_d\lambda}} =
\frac{1}{2}\frac{\sigma_f}{\sqrt{\sigma_dA k_0^2}} .
\end{equation}
Therefore we can reach the conclusion that in the equilibrium state the energy spectrum of our system in the inertial range is flat.
Formally, it is a power law with exponent $\nu=0$  which is very different from the Kolmogorov-Zakharov exponent $\nu=-1$
found in section \ref{sec:kz_spectra}. Recall that the Kolmogorov-Zakharov in the balanced case was found to be marginally nonlocal
and the common wisdom would suggest that it could be fixed by a log correction. As we now see  this is false: our exact solution
has a completely different exponent and has no log factor.

We also see that our exact solution is nonlocal: it does not just depend on the energy flux but it
contains information about both the sources and the sinks as well as about the initial conditions.

%%%%%%%%%%%%%%%%%%%%%%%%%%%%%%%%%%%%%%%%%%%%%%%%%%%%%%%%%%%%%%%%
%
\subsubsection{Viscous friction.}
%
%%%%%%%%%%%%%%%%%%%%%%%%%%%%%%%%%%%%%%%%%%%%%%%%%%%%%%%%%%%%%%%%

Let us now replace the uniform friction by a viscous dissipation keeping the same one-wave initial condition as before.
Equation (\ref{pseudo_fourier}) becomes:
\begin{equation}
\frac{\partial\mathcal{E}(y;\tau)}{\partial\tau}=2 A \left(\cos(k_0 y)-1\right)\mathcal{E}(y;\tau)+  \sigma_\nu\frac{\partial^2\mathcal{E}(y;\tau)}{\partial y^2}+\sigma_f e^{-\frac{k_f^2 y^2}{2}},
\label{full_visc}
\end{equation}
where $  \sigma_\nu$ denotes now the viscosity coefficient and we have used  the initial conditions (\ref{initial}).

To realise this estimation, we need first to get the expression for the steady state solution.
Let us examine the steady state concentrating,  like before in the uniform friction case, on the scales less than the forcing and the initial scales, which in terms of the
pseudo-physical space variables means $y\ll 1/k_0$ and $y\ll 1/k_f$. Performing the same type of expansion in small $y$ as before, we have
\begin{equation}
\sigma_\nu\frac{d^{2}\mathcal{E}(y;\infty)}{d y^2}-\lambda y^2 \mathcal{E}(y;\infty)+\sigma_f=0.
\end{equation}

\begin{figure}
\centering
\includegraphics[width=6cm]{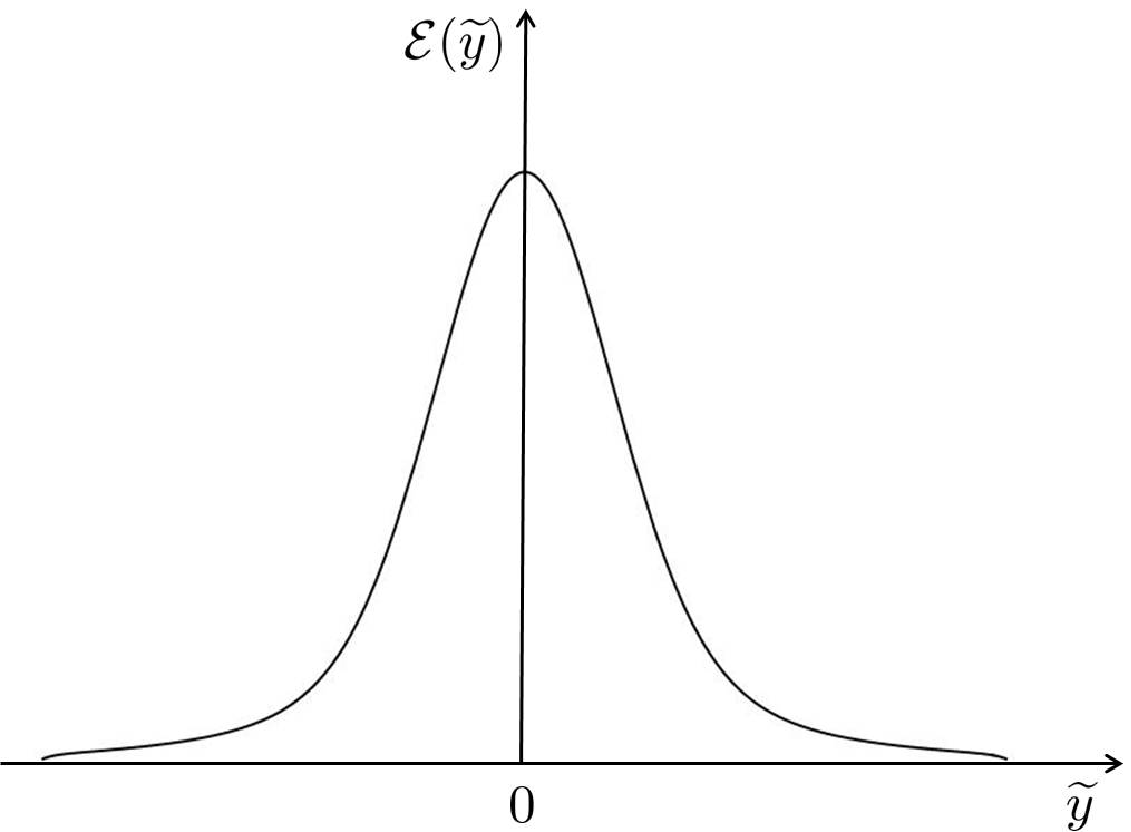}
\caption{\textsf{Stationary solution in the pseudo physical space.\label{fig:pseudo}}}
\end{figure}

By performing the following rescaling
\begin{eqnarray}
\widetilde{y}=y \left(\frac{\lambda}{  \sigma_\nu}\right)^{\frac{1}{4}},\;
\widetilde{\mathcal{E}}=\frac{\sqrt{\lambda   \sigma_\nu}}{\sigma_f}\mathcal{E}
\end{eqnarray}
we obtain
%our \textit{master equation}:
\begin{equation}
\frac{d^2 \widetilde{\mathcal{E}}(\widetilde{y},\infty)}{d \widetilde{y}^2}-\widetilde{y}^2 \,
\widetilde{\mathcal{E}}(\widetilde{y},\infty)+1=0.
\end{equation}
The homogeneous part  here is the equation of the \textit{parabolic cylinder}.
Its solutions are the parabolic cylinder special functions, whose properties and asymptotics
 can be found e.g. in \cite{Abramovich}.
Qualitatively, the behaviour is similar to the one we had in the previous (friction dissipation) example:
it has a maximum at $\widetilde{y}=0$ and it decays
for $\widetilde{y} \to \infty$ (faster than in the previous example).
In  fig. (\ref{fig:pseudo}) we present such a solution in the pseudo-physical space
obtained using \textsf{Matlab}
in the interval with $\widetilde{y}\in\left[-6.1,6.1\right]$. In order to
obtain the decaying solution  we need to take $\widetilde{\mathcal{E}}(0)=1.3110288959$
with high accuracy (to eliminate contribution of  the growing parabolic cylinder function).

Respectively, in the $k_y$-space we again have a flat spectrum in the inertial range,
\begin{equation}
\eta\left(k_y;\infty\right)=C{\sigma_f}\left(\frac{\sigma_\nu}{\lambda^3}\right)^{\frac{1}{4}} \;\;\; \hbox{for}
\;\;\;
k_0,k_f \ll k\ll k_\nu = (\lambda/\sigma_\nu)^{\frac{1}{4}},
\end{equation}
where $C$ is an order one constant.
Once again we see that the spectrum is nonlocal (i.e. it is dependent on the details of the forcing and the sink parameters
rather than just the energy flux), and its exponent is zero (i.e. it is not a log corrected Kolmogorov-Zakharov spectrum).

\begin{figure}
\centering
\includegraphics[width=11cm]{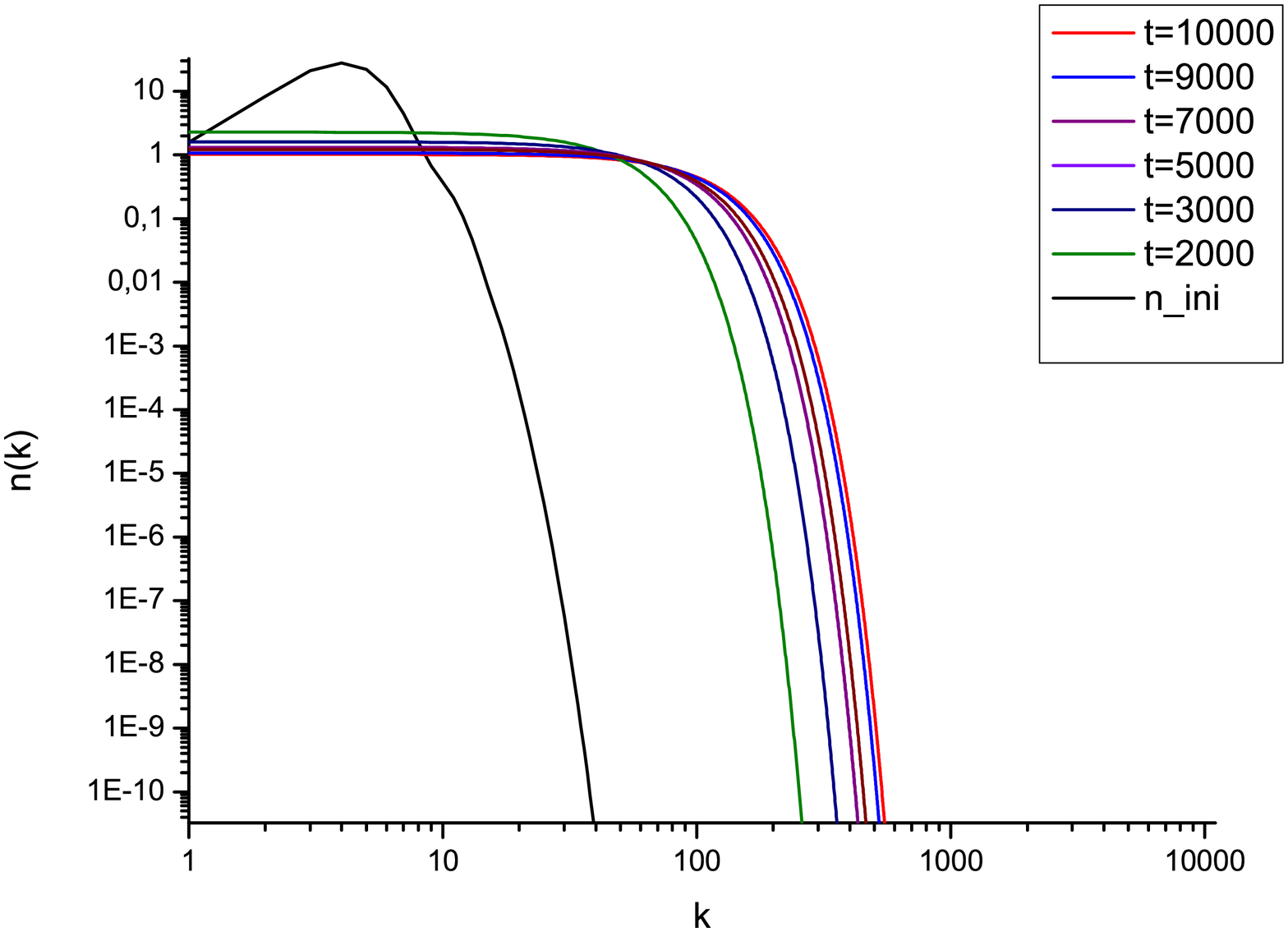}
\caption{\textsf{Spectrum dynamics for a viscous dissipation case $\tau=10000$.\label{fig:tau_10000}}}
\end{figure}

In order to illustrate the dynamical evolution of the spectrum in the viscous dissipation case, we perform direct numerical simulations for the kinetic equation
(\ref{kin_eqn_perp_forced}) in the $k_y$-space.
In this equation we take $\eta^+ = \eta^-, \sigma_d=0$ and ${\mathcal F}_y (k_y)={\mathcal F}_{force}-\sigma_\nu k_y^2$ with
$\sigma_\nu=10^{-6}$.
The result is presented
in
%on the  fig.(\ref{fig:tau_1000}) and the
 Fig. (\ref{fig:tau_10000}).
%\begin{figure}
%\centering
%\includegraphics[width=11cm]{./n_1000.eps}
%caption{\textsf{Spectrum dynamics for a viscous dissipation case $\tau=1000$.\label{fig:tau_1000}}}
%end{figure}
The black curve on this figure represents the initial spectrum with a Gaussian shape
 with a large-scale forcing ${\mathcal F}_{force}=3*10^{-4}/k$ realised for $k\in[3,9]$.
We see that the system converges toward a steady state with a flat spectrum in the inertial range.

%%%%%%%%%%%%%%%%%%%%%%%%%%%%%%%%%%%%%%%%%%%%%%%%%%%%%%%%%%%%%%%%
%
%
\section{Beyond weak turbulence.}
\label{sec:beyond_wt}
%
%
%%%%%%%%%%%%%%%%%%%%%%%%%%%%%%%%%%%%%%%%%%%%%%%%%%%%%%%%%%%%%%%%
 In this section we are providing, at a  qualitative level of rigor, a description for the
energy spectrum behaviour beyond the weak turbulence regime at the late stage of the evolution.
For the wave-kinetic equation to be valid the nonlinearity has to remain weak, i.e. the nonlinear time scale should be much longer than the linear wave period.
This results in the following condition,
\begin{equation}
\frac{t_{nl}}{t_L}=\frac{b_{y} k_{y}}{B_0 k_x}<1.
\end{equation}
Here we have used the estimate for the nonlinear  evolution time taking the standard hydrodynamic non-linear time scale. In our dynamical equations, the nonlinearity related to the magnetic field is of the same order.
Finally, the applicability condition can be rewritten as a limitation for the parallel wave number:
\begin{equation}
k_x>k^*_{x}=\frac{b_{y}k_y}{B_0}.
\end{equation}
It means that for small values of parallel wave vector component near $k_{x}^*$ the kinetic equation  becomes invalid.
 The question about applicability of the wave-kinetic equation near $k_{||}=0$ has been frequently  discussed in the literature, eg. in the 3D
MHD turbulence context \cite{Galtier_2000}.
In particular it was speculated in \cite{Galtier_2000} that  a
sufficient spectrum smoothness for  wavenumbers near $k_{||}=0$ must be present for
the wave-kinetic equation to be applicable.
In the 2D case, the smoothness of spectrum near $k_x=0$ is asymptotically (in time) broken due to a special structure of the kinetic equation.

\begin{figure}
\centering
\includegraphics[width=10cm]{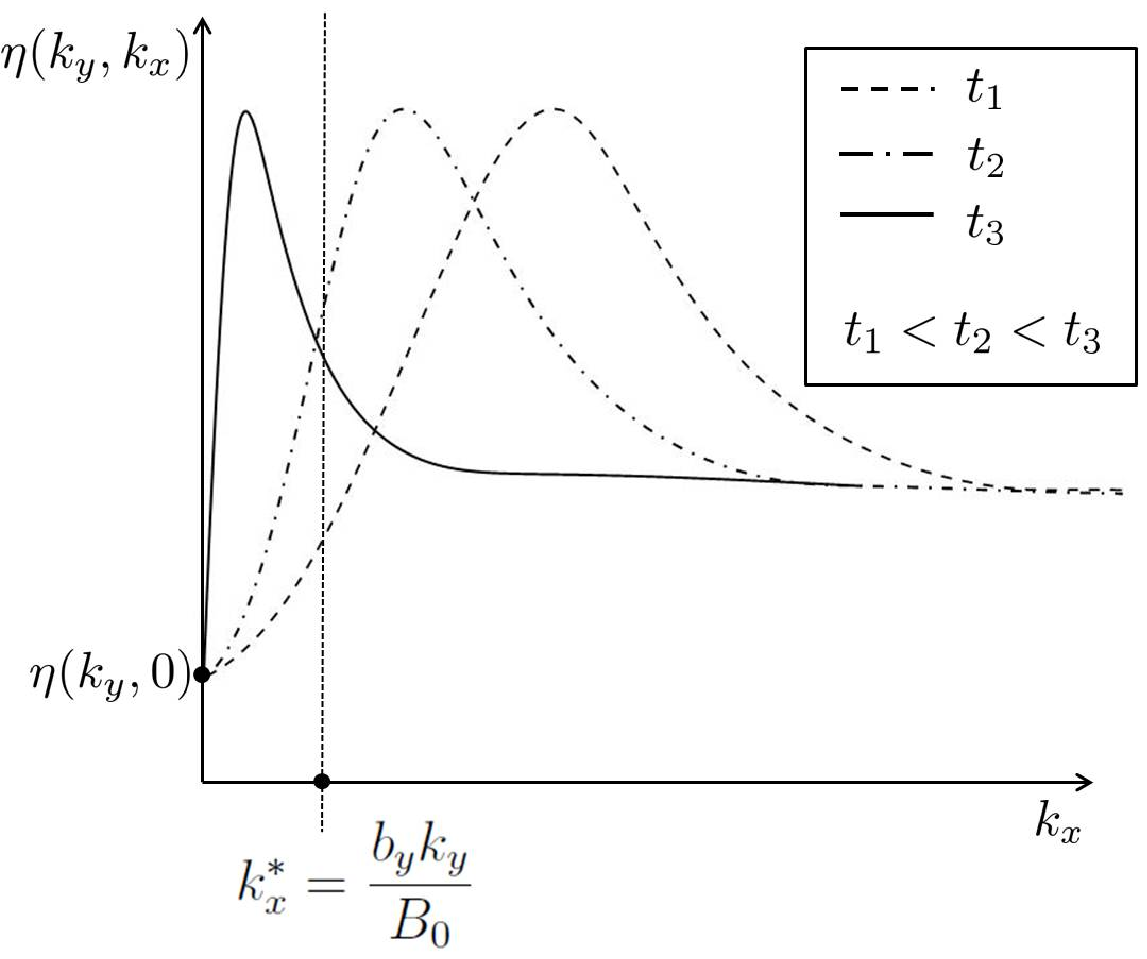}
\caption{\textsf{Spectrum narrowing for large time scales.\label{fig:time_narrowing}}}
\end{figure}

Indeed, as we mentioned before, the wave-kinetic equation for the 2D MHD system
%contains
is formulated in terms of a self-similar "time" variable $\tau=k_{x}^2 t$. Therefore dependence on the parallel wavenumber is still present in the  perpendicular part of the energy spectrum $\eta(k_y,\tau)$ in an implicit way, via  $\tau$.
Such a self-similar dependence on $k_{x}$ is manifested, at each fixed $k_{y}$, in shrinking of the original $k_{x}$ profile along the $k_{x}$-axis as time grows
(see Fig. \ref{fig:time_narrowing}).
The spectrum is narrowing and its derivative is growing near small values of $k_x$, and when it is so steep that a significant variation
occurs over the range $\sim k_{x}^*$,   the kinetic equation breaks down. Time estimate for such a breakdown is
$t\sim (k_x^*)^{-2}$.

Therefore, the weak turbulence description will break down at the late evolution stages, and the wave-kinetic equation will no longer work.
However, it is possible to amend this description to take into account the  strongly nonlinear effects and develop a qualitative theory of subsequent evolution
leading to a steady state.
Below we will present a qualitative argument which will allow us to obtain such a theory.

First of all we note that the three-wave interaction is never exactly resonant: it involves all the quasi-resonant
within a certain small distance from the exact resonant frequency -- so called nonlinear resonance
 broadening $\Gamma\sim t_{nl}^ {-1}$.
In the other words, the delta function in kinetic equation $\delta(2 k_{1x})=\delta\left(\omega_k-\omega_{1}-\omega_2\right)$ should be replaced by a peaked function
$f(k_{1x})$ with a small but finite width $\Gamma$.
For sufficiently smooth spectra the difference from the delta-function can be ignored, but for sharp and narrow
spectra the integrand in the kinetic equation's integral become itself peaked and the delta-function broadening
becomes important.
Its main effect is that of a filter $f(k_{x})$ in the  $k_{x}$ variable which acts to smoothen any sharp changes
over the range $\Delta k \sim  k_x^*$. The energy is no longer conserved separately at each fixed  $k_{x}$.

\begin{figure}
\centering
\includegraphics[width=10cm]{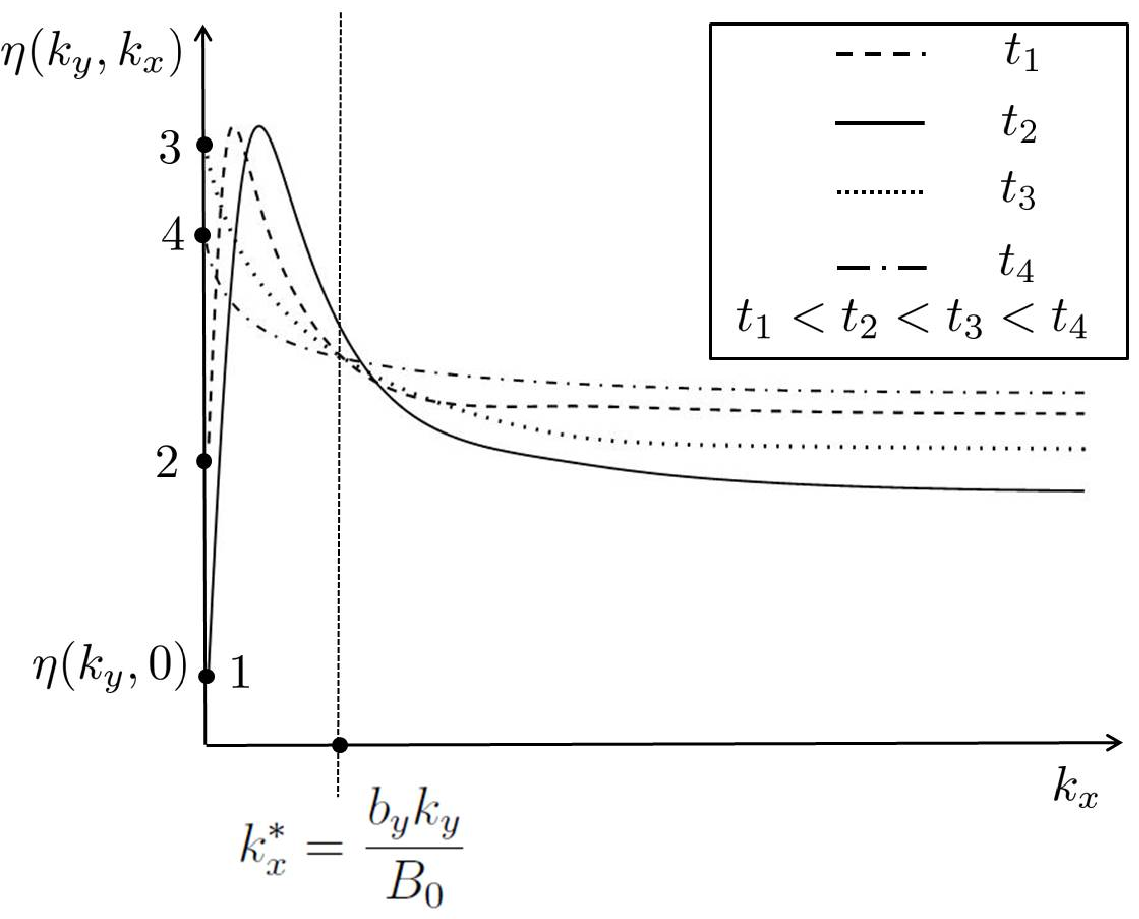}
\caption{\textsf{Gradients smoothing process.\label{fig:grad smoothing}
Four iterations for spectrum value stabilisation are presented on this figure. At the first stage, the gradient of spectrum in the vicinity of the  $k_x=0$ (see point 1) is positive, the initial value of spectrum $\eta(k_y,0)$ increases and reaches point 2, and then, after crossing the maximum,  it moves to point 3, which corresponds to negative slope of the spectrum. Then, the initial value decreases and arrives at the position $4$. This process will continue until the spectrum stabilises at $\eta(k_y,0)=\eta(k_y,\infty)$.}}
\end{figure}

For large wavenumbers $k_x\gg k_x^*$ (where the spectrum remains slowly varying even when it is steep at $k_x\sim k_x^*$)
the kinetic equation could be easily amended by replacing
 $n^{\mp}(0,k_y,0)=\int n^{\mp}\left(k_{1x},k_{1y},\tau\right) \delta(k_{1x}) dk_{1x}$ with
$\langle n^{\mp} \rangle (k_y,\tau)=\int n^{\mp}\left(k_{1x},k_{1y},\tau\right) f(k_{1x}) d k_{1x}$.
For small  wavenumbers, $k_x\sim k_x^*$, the effect of the resonance broadening is not reduced to such a simple modification of just
one function in the integral.
It is clear that the spectrum  at  $k_x=0$, which was fixed in the wave-kinetic approximation,
will suffer changes caused by the smoothing in the direction determined by spectral slope at small $k_x$:
if the gradient is positive (negative) the value will increase (decrease), as illustrated in Fig. \ref{fig:grad smoothing}.
The details of the evolution at the small wavenumbers are not important because the combined action of the
self-similar shrinking and smoothing will lead to a very rapid wipeout of all the gradients in $k_x$
and formation of a steady state with $\eta$ independent of $k_x$.
Correspondingly, the values of  $\eta$  at $k_x=0$ will adjust themselves to the values at  $k_x=\infty$.
After this moment, when the rapid dependencies on  $k_x$ disappear, the kinetic equation in its usual
weak turbulence form is valid once again, and it can be used for finding the
final steady state spectrum. Since $\eta$ is now independent of $k_x$,
the steady state could be readily obtained from the formal condition
$\eta(k_y,0)=\eta(k_y,\infty)$ which simply means our solution
independent of $\tau$; it has nothing to do with the initial/final values of the spectrum
in time or $k_x$.

Thus, the evolution can be summarized as follows. At the early stage,  $t\ll (k_x^*)^{-2}$, the
evolution is described by the three-wave kinetic equation. Then at the advanced stage, with characteristic times scales $t\sim (k_x^*)^{-2}$,
the kinetic equation is broken down by its own evolution. Smoothing of strong gradients in $k_x$ occurs, which
results in spectrum stabilisation and re-emergence of the kinetic equation description at the large time scales $t\gg (k_x^*)^{-2}$. This kinetic equation  describes spectrum evolution within the steady state regime. Let us now consider the properties of such a steady state.

%%%%%%%%%%%%%%%%%%%%%%%%%%%%%%%%%%%%%%%%%%%%%%%%%%%%%%%%%%%%%%%%
%
\subsection{Spectrum evolution in the steady-state}
%
%%%%%%%%%%%%%%%%%%%%%%%%%%%%%%%%%%%%%%%%%%%%%%%%%%%%%%%%%%%%%%%
Let us  now analyse the steady state.
Based on what was said in the end of the previous section, we
will seek a  $\tau$-independent solution of the pseudo-Fourier space equation (\ref{pseudo_fourier}):
\begin{equation}
\mathcal E^2 (y)-\mathcal E(y)\left(\mathcal E(0)+\sigma_d\right)+\widehat{{\mathcal F}}(y)=0
\end{equation}
(formally coinciding with the condition  $\mathcal{E}(y,0)=\mathcal{E}(y,\infty) = \mathcal{E}(y)$).
Considering this equation at $y=0$ we have
$\mathcal E(0) =  \widehat{{\mathcal F}}(0)/\sigma_d$.
Also we have $\mathcal{E}(0)=\int \eta(k_y)d k_y > 0$ (because
$\eta(k_y) \ge 0$).
Solving the quadratic equation, we have:
\begin{equation}
\mathcal{E}(y)=\frac 1{2}
{(\sigma_d  + \widehat{{\mathcal F}}(0)/\sigma_d)} \pm
\frac 1{2} \left((\sigma_d +\widehat{{\mathcal F}}(0)/\sigma_d)^2-4\widehat{{\mathcal F}}(y)\right)^{1/2}.
\label{st_sp}
\end{equation}
To satisfy condition $\mathcal E(0) =  \widehat{{\mathcal F}}(0)/\sigma_d$,
we must choose ``+" if $\widehat{{\mathcal F}}(0) >\sigma_d^2$ and ``-"
otherwise.

We suppose that the forcing  decays at infinity, $\lim_{y\rightarrow\infty}\widehat{{\mathcal F}}(y)\rightarrow 0$,
which is the case eg. for the Gaussian forcing.
This means that we do not force the $k_y=0$ mode.
%An important point that we need to understand about this equilibrium solutions, is about its behaviour for $y\rightarrow\infty$, which is equivalent on the physical space to $k\rightarrow 0$.
Then we see that
$\lim_{y\rightarrow\infty} \mathcal{E}(y) \to 0$
if $\widehat{{\mathcal F}}(0) <\sigma_d^2$ and
$\lim_{y\rightarrow\infty} \mathcal{E}(y) \to \sigma_d  + \widehat{{\mathcal F}}(0)/\sigma_d$
if $\widehat{{\mathcal F}}(0) > \sigma_d^2$ .
 In the second case we
have a spectrum with a delta function at $k_y=0$. Thus we observe an interesting phenomenon
of condensation into $k_y=0$ mode in the cases when
the forcing prevails over the dissipation at the small scales $y$ (corresponding
to high $k_y$'s). In the first case $\mathcal{E}(y) $ is a monotonously decreasing (to 0) function of $y$, where as in the second case it is
monotonously increasing (asymptoting to constant).

The first case is physically more relevant, because in most cases
of interest dissipation dominates over forcing at the small scales.
In this case $\mathcal{E}(y) $ behaves qualitatively similar as in
the two examples considered in section \ref{sec:kin_eqn_dyn}.
Namely, if we take the same Gaussian forcing as in these two examples,
we will have $\mathcal{E}(y) $ which has a maximum at $y=0$, smooth
everywhere (including $y=0$) and rapidly decaying to zero for $y\to\infty$.
However, there is a big difference from the previous examples in that
now the characteristic width of function $\mathcal{E}(y) $, and
respectively the width of the spectrum in the $k_y$ variable, is of the
same order as the width of the forcing function.
Therefore, there is no inertial range in the final steady state considered here.
This is an even stronger case of the nonlocal interaction than
in the two examples considered before. Both the forcing and the dissipation parameters enter in the final answer, but not the parameters of the initial condition: the steady state beyond the weak turbulence has already
forgotten all the initial data.

%%%%%%%%%%%%%%%%%%%%%%%%%%%%%%%%%%%%%%%%%%%%%%%%%%%%%%%%%%%%%%%%
%
%
\section{Summary.}
%
%
%%%%%%%%%%%%%%%%%%%%%%%%%%%%%%%%%%%%%%%%%%%%%%%%%%%%%%%%%%%%%%%%%
\label{sec:conclusions}

We have shown that the three-wave interactions for the pseudo-Alfv\'en  waves (PAW)  in the 2D MHD system are non-empty and it is possible to obtain a three-wave kinetic equation within the weak turbulence approach. These interactions take place in the second order of the anisotropy parameter.
%The implementation of the general perturbative procedure leading to such an equation is given in the \ref{app:WT_procedure}.

We found Kolmogorov-Zakharov power law spectra for PAW in 2D MHD system and showed that they
are not realisable due to divergence of the collision integrals of the kinetic equation. In the balanced case this divergence is marginal. This is an indirect indication that the 2D PAW turbulence is nonlocal: it is dominated by interaction of waves with very different in size wavelengths. Our full analytical solution of the kinetic equation confirms such a nonlocality. It also dispels the myth that all marginally nonlocal spectra can be ``fixed" by a logarithmic correction.

The crucial technique for our analysis is passing to pseudo-physical space
{\it via} Fourier-transforming the kinetic equation and using a self-similar effective time variable. This has allowed us to dramatically
simplify the kinetic equation, solve it analytically in some important
cases and to fully analyse in the other important cases.
The two main examples we analysed have a Gaussian-shaped forcing of low
wavenumbers and a dissipation represented by either uniform friction or by viscosity. The first case is solvable analytically, and the second one is shown to possess a similar behaviour. Namely, the spectrum evolves independently
at each $k_x$ and it tends to a flat steady-state spectrum in the inertial range
(which is not a log-corrected Kolmogorov-Zakharov spectrum).

At each fixed $k_y$, the spectrum develops sharp gradients in $k_x$ at small
$k_x$, which eventually leads to the breakdown of the weak turbulence description. We present a qualitative argument about what follows after this
moment. We argue that the effect of strong turbulence is to smoothen the
sharp gradients via the nonlinear resonance broadening effect. This will lead to a steady state with no gradient in $k_x$ for which the weak turbulence kinetic equation formally works once again, and we present an analytical
solution for such a steady state.

On the practical side, one should derive from our work a warning that the 2D and the 3D MHD systems are dramatically different, and one should be careful
when extrapolating the 2D results, eg. numerical ones, onto the 3D case.
Indeed, in contrast to 2D, in 3D there is no gradient sharpening at small parallel wave numbers, and the Kolmogorov-Zakharov spectrum is a local and well behaved solution.

%%%%%%%%%%%%%%%%%%%%%%%%%%%%%%%%%%%%%%%%%%%%%%%%%%%%
\section{Acknowledgement}
Sergey Nazarenko gratefully acknowledges support from the government of Russian Federation  via grant
No. 12.740.11.1430 for supporting research of teams working under supervision of
invited scientists.

%%%%%%%%%%%%%%%%%%%%%%%%%%%%%%%%%%%%%%%%%%%%%%%%%%%%

%
\appendix
\section{Derivation of the wave-kinetic equation }
\label{app:WT_procedure}
%
%
%%%%%%%%%%%%%%%%%%%%%%%%%%%%%%%%%%%%%%%%%%%%%%%%%%%%%
In  section \ref{sec:alfven_waves}
we wrote the 2D MHD system in the interaction representation (\ref{eq:mhd_interact}),
which comprises a starting point

for derivation of the wave-kinetic equation.
Let us define the wave spectrum as
$$
n_k^{\pm} = L^2 \epsilon^2 \langle |c_k^{\pm}|^2  \rangle,
$$
where the average is taken over the random initial conditions,
and $L^2$ is the area of the periodic box. With this normalisation $n_k$ tends to a finite limit $\sim \epsilon^2$ as $L\to\infty$ provided that the wave density is finite and uniform in the 2D physical space.

The next step consists in making use of the time scales separation.
We are introducing an intermediate  time scale $T$ which should be much smaller than the typical non-linear time, $t_{nl} = 2\pi/(\epsilon^2\omega)$,
  and much greater than the linear wave period, $t_L=2\pi/\omega$.
Taking $T=2\pi/(\epsilon\omega)$ will satisfy these conditions,
$t_L\ll T\ll t_{nl}$.
Then we are looking for solutions at time  $t=T$ in the form series in  small $\epsilon$:
\begin{equation}
c_k^{\pm}(T)=c_k^{(\pm,0)}+\epsilon c_k^{(\pm,1)}+\epsilon^2 c_k^{(\pm,2)}+\dots \, ,
\label{expa}
\end{equation}
where we suppose that the lowest order amplitudes
$c_k^{\pm,(0)}=c_k^{\pm}(0)$ correspond to the linear regime.

For the spectrum we have:
\begin{equation}
[n_k^{\pm}(T)-n_k^{\pm}(0)]/(\epsilon^4 L^2)=\left\langle\left|c_k^{(\pm,1)}\right|^2\right\rangle
+\left\langle c_k^{(\pm,0)*}c_k^{(\pm,2)}\right\rangle
+\left\langle c_k^{(\pm,0)}c_k^{(\pm,2)*}\right\rangle ,
\label{three_wave}
\end{equation}

After substituting  expansion (\ref{expa}) into  equation (\ref{eq:mhd_interact})
in  the first order we have:
\begin{equation}
c_k^{(\pm,1)}(T)=\sum_{1,2} V_{k,1,2} \Delta_T\left(\pm 2 k_{1x}\right)\ c_1^{(\mp,0)}c_2^{(\pm,0)}\delta_{12}^k,
\label{1_order_ampl}
\end{equation}
where
\begin{equation}
\Delta_T\left(\pm2 k_{1x}\right)= \int_0^T e^{\pm 2 i k_{1x}t}\  d t=\frac{e^{\pm i 2 k_{1x} T}-1}{\pm 2 i k_{1x}}.
\end{equation}
For the second order  we can write:
\begin{eqnarray}
{c}_k^{(\pm, 2)}=\sum_{1,2,3,4} V_{k,1,2} \, \delta_{12}^k
\left[ V_{2,3,4}\delta_{34}^2 c_1^{(\mp,0)} c_3^{(\mp,0)} c_4^{(\pm,0)} E\left(\pm 2k_{1x},\pm  2k_{3x}\right)\right.\\
\nonumber
\left.+ V_{1,3,4}\delta_{34}^1 c_2^{(\pm,0)} c_3^{(\pm,0)} c_4^{(\mp,0)} E\left(\pm 2k_{1x},\mp  2k_{3x}\right) \right],
\end{eqnarray}
with
\begin{equation}
E(x,y)=\int_0^T e^{i xt} \Delta_t\left(y\right) d t.
\end{equation}

Next, we are going to assume that the initial amplitudes
${c}_k^{(\pm, 0)}$ are Gaussian random variables which are statistically independent at each $\bf k$, and use Wick's rule:
\begin{equation}
\left\langle
c_1^{{(\pm,0)}}c_2^{{(\pm,0)}} c_3^{{(\mp,0)}}c_4^{{(\mp,0)}}
\right\rangle
=\delta({\bf k}_1 +{\bf k}_2)
\delta({\bf k}_3 +{\bf k}_4)
 \langle |c_1^{{(\pm,0)}}|^2  \rangle \langle |c_3^{{(\mp,0)}}|^2  \rangle
\, .
\end{equation}
We also remember that because the physical space amplitudes are real functions we have $\left({c}^{(\pm, 0)} ({\bf k})\right)^* =
{c}^{(\pm, 0)} (-{\bf k})$.

We have:
%$\Delta_T^*(\pm 2 k_{1x})=\Delta_T(\mp 2 k_{1x})$ and we proceed with triadic interaction calculation:
\begin{eqnarray}
\left\langle\left|c_k^{(\pm,1)}\right|^2\right\rangle&=&
 \sum_{1,2,3,4}V_{k,1,2} V_{k,3,4} \Delta_T\left(\pm 2k_{1x}\right) \Delta_T^*\left(\pm 2k_{3x}\right)
\nonumber
\\
&\times&
\left\langle c_1^{(\mp,0)} c_2^{(\pm,0)} \left(c_3^{(\mp,0)}\right)^* \left(c_4^{(\pm,0)}\right)^*
\right\rangle \delta^k_{12}\delta^k_{34}
\label{prop_1_11}
\\
&=&
\frac 1{L^4\epsilon^4} \sum_{12} \left|V_{k,1,2}\right|^2 \left|\Delta_T\left(\pm 2k_{1x}\right)\right|^2 n_1^{\mp}n_2^{\pm} \delta^k_{12},
\nonumber
\end{eqnarray}
and
\begin{eqnarray}
%\fl
\left\langle\left(c_k^{(\pm,0)}\right)^*c_k^{(\pm,2)}\right\rangle&=&
 \sum_{1234} V_{k,1,2} \delta_{12}^k\left[V_{2,3,4}\delta_{34}^2 E\left(\pm 2 k_{1x},\pm 2 k_{3x}\right)
\left\langle \left(c_k^{(\pm,0)}\right)^* c_1^{\mp,0}c_3^{\mp,0}c_4^{\pm,0}\right\rangle\right]\nonumber\\
&=&- \frac 1{L^4\epsilon^4} \sum_{12} \left|V_{k,1,2}\right|^2 E\left(\pm 2 k_{1x},\mp 2k_{1x}\right) n_1^{\mp}n_k^{\pm} \delta_{12}^k ,
\label{prop_0_2}
\end{eqnarray}
where we have used  abbreviations $n_1^{\mp}=n^{\mp}({\bf k}_1,t)$, $n_2^{\mp}=n^{\mp}({\bf k}_2,t)$ and $\delta_{12}^k =
\delta({\bf k}_1 +{\bf k}_2 - {\bf k})$.

Next we note that
\begin{equation}
\Im E(\pm 2k_{1x},\mp 2k_{1x})=-\Im E(\mp 2 k_{1x},\pm 2 k_{1x})
\end{equation}
and
\begin{equation}
\Re E(\pm 2k_{1x},\mp 2k_{1x})=\Re E(\mp 2 k_{1x},\pm 2 k_{1x})= \frac{\sin^2\left(k_{1x}T\right)}{2\ k_{1x}^2}.
\end{equation}
Let substitute expressions  (\ref{prop_1_11}) and (\ref{prop_0_2})  into the eq. (\ref{three_wave}),
\begin{equation}
n_k^{\pm}(T)-n_k^{\pm}(0)= \frac 1{L^2}  \sum_{1, 2} \left|V_{k,1,2}\right|^2 n_1^{\mp}\left(n_2^{\pm}-n_k^{\pm}\right) \delta_{12}^k
\frac{\sin^2\left(k_{1 x} T\right)}{k_{1 x}^2},
\label{three_wave_sum}
\end{equation}
where we have used that $\left|\Delta_T(2 k_{1x})\right|^2=\sin^2(k_{1x}T)/k_{1x}^2$.

Now we take the infinite box limit, $L\to \infty$, and pass to the continuous
description in the ${\bf k}$-space using the rule
$$
\frac 1{L^2}  \sum_{1, 2}  \delta_{12}^k \to
\int \delta_{12}^k \, d{\bf k}_1 d{\bf k}_2,
$$
where $\delta_{12}^k $ in the integrand means Dirac's delta (recall that
it is Kronecker delta in the sum).

At the next stage of the wave-kinetic procedure we need to use the weakness of the non-linearity in our system by taking the limit $\epsilon\rightarrow 0$, which is equivalent to $T\rightarrow\infty$.
For the r.h.s. of the eq. (\ref{three_wave_sum}), we obtain:
\begin{equation}
\lim_{T\rightarrow\infty}\frac{\sin^2(k_{1x}T)}{k_{1x}^2}=\pi T \delta\left(k_{1x}\right).
\end{equation}
Then after multiplying both parts of the eq.(\ref{three_wave_sum}) by $1/T$,
its l.h.s. becomes:
\begin{equation}
\frac{n(T)-n(0)}{T} \to\dot n (T),
\end{equation}
where we took into account that time $T$ is much less than the nonlinear time at which the spectrum evolves.

After these steps we can finally write down  the kinetic equation:
%Finally we obtain the expression for the kinetic equation, which results from the triadic waves interactions:
\begin{equation}
\dot{n}_k^{\pm}=\pi \int V_{k 12}^2 n_1^{\mp}\left[ n_2^{\pm}- n_k^{\pm}\right]
\delta\left(\mathbf{k}-\mathbf{k}_1-
\mathbf{k}_2\right)\delta(2\ k_{1x}) d\mathbf{k}_1\ d\mathbf{k}_2 \, .
\end{equation}
%%%%%%%%%%%%%%%%%%%%%%%%%%%%%%%%%%%%%%%%%%%%%%%%%%%%%%%%%%%%%%%%%
%\section{Recovery of the wave-kinetic equation in two-dimensions from the three-dimensional case.}
%\label{app: 3D_2D}
\section{Locality study for Kolmogorov-Zakharov solutions.}
\label{app:convergence}
In order to explore realisability of Kolmogorov-Zakharov spectra,  $ \eta(k_y;\infty)^{\pm}\propto k_y^{\nu^{\pm}}$, we need to proceed with a convergence study of the collisional integrals:
\begin{equation}
\int_{-\infty}^{\infty} \delta\left(k_{1y}+k_{2y}-k_y\right)
|k_{1y}|^{\alpha_{\pm}}\left(|k_{2y}|^{\alpha_{\mp}}-|k_y|^{\alpha_{\mp}}\right) d k_{1y}d k_{2y}
\label{int_conv_1}
\end{equation}

Let us consider the first one, choosing $\alpha_+$ at the exponent of $|k_{1 y}|$.
There are three singular points:
\begin{enumerate}
\item {$k_{1y},k_{2y}\rightarrow\infty$},
\item{$k_{1y}\rightarrow 0, k_{2y} \to k_y$},
\item{$k_{2y}\rightarrow 0,  k_{1y} \to k_y$}.
\end{enumerate}

At the first point, we should use the fact that  the integral
$\int_{1}^{\infty}|x|^{\nu} d x$ converges when $\nu<-1$.
After substituting $k_{2y}=k_y-k_{1y}$, we have :
$$
\int_1^{\infty} |k_{1y}|^{\alpha_+}\left(|k_y-k_{1y}|^{\alpha_{-}}-
|k_y|^{\alpha_-}\right) dk_{1 y}.
$$
Then two cases are possible:

\begin{itemize}
\item when $\alpha_->0$, the main contribution is made by:
$$
\int_1^{\infty} |k_{1y}|^{\alpha_++\alpha_-} dk_{1 y},
$$
which is convergent for $\alpha_++\alpha_-<-1$,

\item when $\alpha_-<0$ the expression for the main contribution is made by:
$$
\int_1^{\infty} |k_{1y}|^{\alpha_+}|k_y|^{\alpha_-} dk_{1 y},
$$
and is convergent for $\alpha_+<-1$.

\end{itemize}
%The second integral converges for $\alpha_{+}<-1, \forall \alpha_-$.
At the second singular point, after integration out $k_{2y}$ using the $\delta$-function in (\ref{int_conv_1}), we have:
\begin{equation}
\int_0^{\epsilon} |k_{1y}|^{\alpha_+}\left(|k_y-k_{1y}|^{\alpha_-}-|k_y|^{\alpha_-}\right) dk_{1 y}\sim
\int_0^{\epsilon} k_{1y}^{\alpha_++1} dk_{1 y},
\end{equation}
and we get the convergence condition $\alpha_+>-2$.
To get this condition, we have performed the series expansion:
$|k_y-k_{1y}|^{\alpha_-}=|k_y|^{\alpha_-} (1+
\alpha_-k_{1y}/k_y ) +\dots$, and we have used the fact that the integral $\int_0^{\epsilon} x^{\nu} dx$ is convergent for $\nu>-1$.

To obtain the convergence condition for the last singular point, we integrate out
$k_{1y}$ using the $\delta$-function:
\begin{eqnarray}
&&\int_0^{\epsilon} |k_{y}-k_{2y}|^{\alpha_+}\left(|k_{2y}|^{\alpha_-}-|k_y|^{\alpha_-}\right) dk_{2 y}\sim
\\
\nonumber
&&\int_0^{\epsilon} |k_{y}|^{\alpha_+}|k_{2y}|^{\alpha_-} dk_{2 y}-\int_0^{\epsilon}|k_y|^{\alpha_++\alpha_-} dk_{2 y} \, .
\end{eqnarray}
The second integral is always convergent, and
the first one is convergent for $\alpha_->-1$.

\begin{figure}[h!]
\centering
\includegraphics[width=10cm]{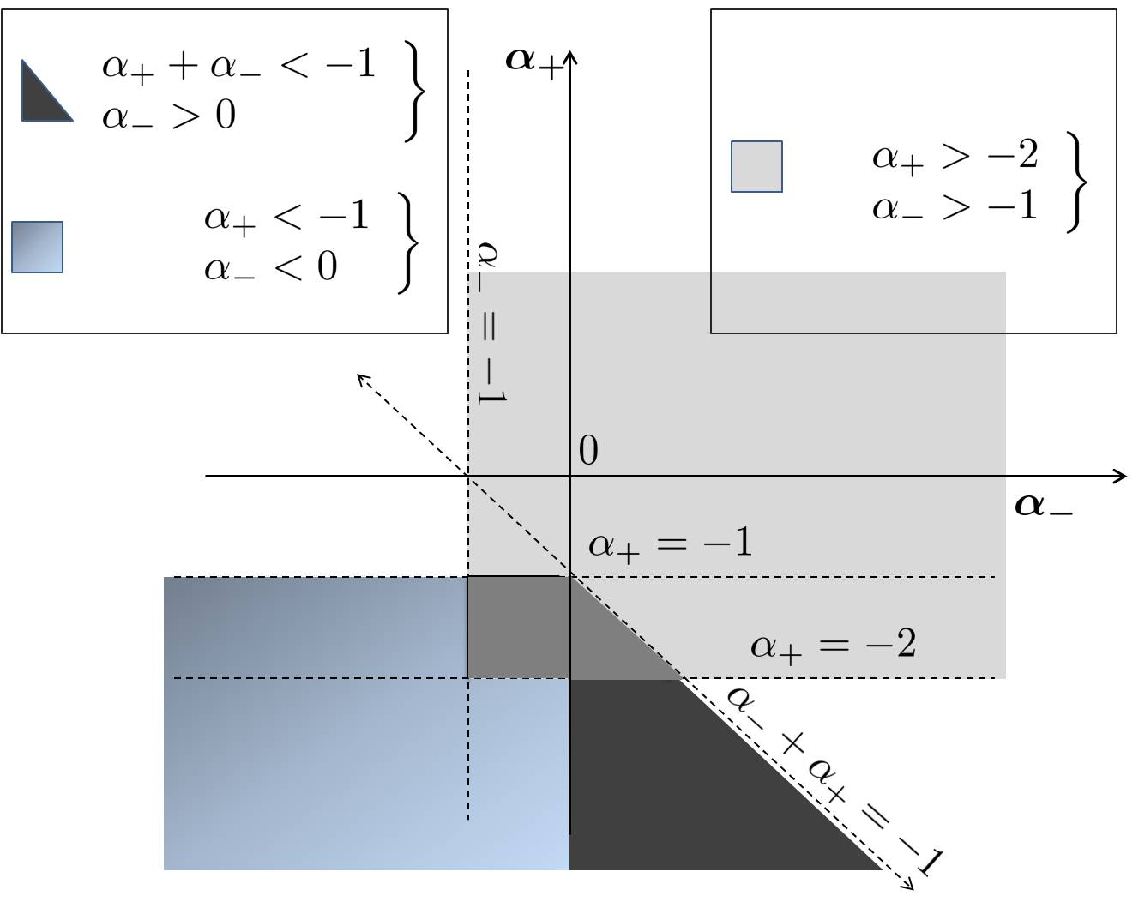}
\caption{\textsf{Locality study for Kolmogorov-Zakharov spectrum. \label{fig:convergence_1}}}
\end{figure}

Finally,  the convergence region  for the first collisional integral (\ref{int_conv_1}) in the space of indices is:
\begin{eqnarray}
%\fl
     \left\{    \left\{ (\alpha_++\alpha_-<-1)  \cap
           (\alpha_->0)
     \right\}  \cup \left\{
         (\alpha_+<-1)  \cap
        (\alpha_-<-0)
     \right\}    \right\}
      \cap
(\alpha_+>-2)  \cap
(\alpha_->-1) \, .
\nonumber
\end{eqnarray}
It is represented  by the grey trapezoid in Fig. \ref{fig:convergence_1}.

 To find the convergence zone for the second integral of (\ref{int_conv_1}) (with $\alpha_-$ in the exponent of $k_{1y}$) we  just take reflection of the convergence zone for the first integral with respect to the line $\alpha_{-}=\alpha_{+}$. Finally, to get the convergence conditions for both collisional integrals one should take the intersection of the both zones.

As we can see in Fig. \ref{fig:convergence_1}
such an intersection produces a zero set. There are no power law exponents
$\alpha^{-}$ and $\alpha^{+}$ for which both collision integrals would be convergent, and there is a single point which corresponds to
marginal (logarithmic) divergence, $\alpha^{-}=\alpha^{+}=-1$. This point corresponds to the Kolmogorov-Zakharov spectrum in the balanced turbulence case. Common wisdom \cite{Kraichnan_1970} is that such marginally nonlocal spectra can be fixed
by a logarithmic correction. However, in the main text of this paper we show that this is not the case.

\bibliographystyle{apsrev4-1}
\bibliography{testbib}

\end{document}